\documentclass[preprint,aps,prc,amsmath,amssymb,showpacs,showkeys,floatfix,diagbox]{revtex4}

\usepackage{graphicx}
\usepackage{bm}
\usepackage{epstopdf}
\usepackage{float}
\usepackage{booktabs}
\usepackage{dcolumn}
\usepackage{bm}
\usepackage{mathrsfs}
\usepackage{amsmath}
\usepackage{footnote}
\usepackage{accents}
\usepackage{mathtools}
\usepackage{tensor}
\usepackage[titletoc]{appendix}
\usepackage{color}
\usepackage{accents}

\begin{document}



\title{Cosmologies in $f(R,\mathcal{L}_{m})$ theory with non-minimal coupling between geometry and matter}
\author{Sergio Bravo Medina}
\email{sergiobravom@javeriana.edu.co}
\affiliation{
Departamento de F\'isica,\\ Pontificia Universidad Javeriana, Cra.7
No.40-62, Bogot\'a, Colombia
}
\author{Marek Nowakowski}
\email{marek.nowakowski@ictp-saifr.org}
\affiliation{
 ICTP-South American Institute for Fundamental Research,
Rua Dr. Bento Teobaldo Ferraz 271, 01140-070 S\~ao Paulo, SP Brazil
}
\author{Ronaldo V. Lobato}
\email{contact@rvlobato.com}
\affiliation{
  ICRANet, Piazza della Repubblica 10, Pescara, 65122, Italy
}

\author{Davide Batic}
\email{davide.batic@ku.ac.ae}
\affiliation{
Department of Mathematics, \\
Khalifa University of Science and Technology,
Sas Al Nakhl Campus,
P.O. Box 2533 Abu Dhabi,
United Arab Emirates
}

\date{\today}

\begin{abstract}
Among the recent extensions to standard General Relativity, $f(R,\mathcal{L}_{m})$ gravity has risen an interest given the possibility of coupling between geometry and matter.
We examine the simplest model with non-minimal coupling in the context
of cosmology. We pay special attention to the question of how far this
model could reproduce the observational fact of our universe.

\end{abstract}

\pacs{ }
\maketitle

\section{Introduction}
Since the discovery of the accelerated expansion of the universe, cosmologists have developed the $\Lambda CDM$ model, a standard cosmological model rooted in Einstein's general relativity \cite{Einstein}. This model incorporates a positive cosmological constant $\Lambda$ \cite{Lambda}, the Friedmann-Robertson-Walker metric \cite{FRW}, and the concept of Dark Matter (DM) \cite{DM}. The otherwise robust model is not completely without problems.
To start with, despite numerous efforts, a candidate for DM has not yet been found. The
second problem is coined as Hubble tension, careful phrasing of the fact that different measurements of the Hubble constant yield different results \cite{HubbleTension}. In the future, as we seek to explain these issues, we may require a new theoretical model. This could involve modifying Einstein's gravity \cite{Modifications}, replacing the cosmological constant with different models of Dark Energy (DE) \cite{DE}, and exploring alternative candidates for DM \cite{Alternatives}. The number of extensions of general relativity as well as the number of DE models is, of course, quite large \cite{Sotiriou, DeFelice, Olmo,
    HLNO, Myrzakulov, JWu, Katirci, Roshan, Board, Cai, RYang,
    Capozziello, Heisenberg, Khyllep, Koussour, Jimenez,
    Guangjie,
    Shiravand, XHarko, Bahamonde, Obukhov, HeisenbergKuhn}.

Despite the remarkable success of the $\Lambda$ Cold Dark Matter ($\Lambda$CDM) model in explaining a wide array of cosmological observations, from the cosmic microwave background (CMB) anisotropies to the large-scale structure of the universe, several significant challenges remain unresolved. The nature of dark energy, epitomized by the cosmological constant $\Lambda$, poses a profound theoretical conundrum, known as the "cosmological constant problem" \cite{Lambda}. The astonishingly small value of $\Lambda$, as required to explain the observed acceleration of the universe, stands in stark contrast to theoretical predictions from quantum field theory, typically larger by many orders of magnitude \cite{Lambda}.

Moreover, the $\Lambda$CDM model does not escape from the Hubble tension, a growing discrepancy between the values of the Hubble constant, $H_0$, measured directly from local astronomical observations and those inferred from the CMB under the $\Lambda$CDM framework \cite{Verde}. These inconsistencies point to potential physics beyond the standard model, either in the form of new particles or fields, or through modifications to the theory of general relativity itself.

In response to these challenges, a plethora of modified gravity theories have been proposed as alternatives to general relativity, aiming to provide a more comprehensive theoretical framework that can naturally incorporate the phenomena attributed to dark energy and dark matter. Theories such as $f(R)$ gravity, where the Ricci scalar $R$ in the Einstein-Hilbert action is replaced by a function $f(R)$, offer promising ways to explain cosmic acceleration without the need for a cosmological constant \cite{Sotiriou,DeFelice}.

Expanding further, $f(R,T)$ theories introduce a coupling between matter, represented by the trace of the energy-momentum tensor $T$, and geometry, providing a framework to explore the effects of such couplings on the dynamics of the universe \cite{HLNO}. The $f(R,\mathcal{L}_m)$ models we focus on in this paper extend this idea by incorporating the matter Lagrangian $\mathcal{L}_m$ directly into the gravitational action, potentially offering new insights into the interaction between dark matter and dark energy, and their impact on the evolution of cosmic structures \cite{FRLm}.

Our motivation for selecting $f(R,\mathcal{L}_m)$ gravity stems from its ability to unite the geometrical and material sectors of the universe in a single, coherent theoretical framework. This approach not only allows for the exploration of the cosmic acceleration and dark matter problems from a new angle but also provides a platform for testing the limits of Einstein's general relativity on cosmological scales. By investigating the cosmological implications of $f(R,\mathcal{L}_m)$ gravity, we aim to contribute to the ongoing discussion on viable alternatives to $\Lambda$CDM, exploring whether these theories can offer a more satisfactory explanation of observational phenomena without some of the fine-tuning issues that plague the standard model.

It is therefore a priori not clear
which class of models is theoretically preferred over the others.
We think that simplicity of the modification or, in other words,
Occam's razor, could serve us here as guiding principle.
Within a range of models, this possibility seems quite likely. Among the wide class of models based on a Lagrangian function of the type $f(R)$, where $R$ is the Ricci scalar \cite{Odintsov}, one could choose to focus on $R^2$ gravity \cite{fR2} and the resulting cosmological models \cite{fR2cosm}. Recently, another class has been examined, based on
$f(R, \mathcal{L}_{m})$ where $\mathcal{L}_m$ \cite{FRLm, FRLm2,
  FRLm3}
represents the matter
Lagrangian, often taken to be the energy density $\rho$, pressure $p$, or the trace of the
energy-momentum tensor $T$-with all these quantities being diffeomorphically invariant. In this class, an additional term in the Einstein-Hilbert Lagrangian of the form $\sigma R\mathcal{L}_m$ would be considered a mild extension of the previous model. Moreover, it provides a straightforward coupling of the geometry encoded in $R$ with matter represented
astrophysically by $\mathcal{L}_m$.  The
astrophysical and other implications of such a theory have been examined in
\cite{Montelongo, Lobato, Lobato2, Lobato3} with consequences for Neutron star and
White Dwarf physics. Some cosmological aspects with specific choices
of $\mathcal{L}_m$ have been considered in \cite{CosmologyFRLm,
  TransitFRLm, AcceleratingFRLm, ViscousDM, ConstrainFRLm,
  Gonclaves}.
Except for \cite{Gonclaves}, the choice of the function
$f(R, L_m)$ is different from ours. The
$f(R,\mathcal{L}_{m})$ theories have their own theoretical
peculiarities that require careful examination. One notable aspect is
that choosing
$\mathcal{L}_m=\pm \rho, \pm p, T$ does not
lead, through standard metric variation to the energy-momentum tensor of a perfect fluid $T_{\mu \nu}$ \cite{EnergyMomentumHarko,
  MatterLagrangian, MinazzolliHarko, Brown}. On the other hand, a constrained variation
based on the Lagrangian with constraints can result in
the prefect fluid energy-momentum tensor. However, this comes at the cost of  $T_{\mu \nu}$ no longer being conserved.  We show that explicitly using
the new Friedmann equation in $f(R, \mathcal{L}_{m})$.

In solving the new Friedmann equations, we try to make the model resemble our present universe. While it seems possible to achieve this by choosing adequate initial values, the evolution into the past and future can bring some surprises, as will be shown below.  One of the simplest and model independent feature of our
universe is the lower limit on its age as given for example by the
redshift of the oldest galaxies \cite{galaxies}.
A second constraint independent of the model comes from uranium decays  \cite{uranium}.
If a cosmological model, as
appealing as it may be, cannot
reproduce these facts, it is then certainly not a good candidate to
describe our universe. We will pay a special attention to these
questions while examining the details of the model under discussion.

The paper is organized as follows. In Section II, we outline the basics
of $f(R,\mathcal{L}_{m})$ theory with minimal coupling, denoted as $\sigma
R\mathcal{L}_m$. In this section, we also present the new Friedmann equations with a cosmological constant in a dimensionless form suitable for numerical integration.
In Section III, we present the generalization of one of the
Friedmann equations. In its standard form, it reads
$\Omega_{\Lambda,0} +\Omega_{m,0}=1$. However, in the context of $f(R,\mathcal{L}_{m})$ theory, it generalizes to
$\Omega_{\Lambda,0} +\Omega_{m,0}+ \Omega_{\sigma, 0}=1$. In the same section, we discuss the possible initial values, a necessary undertaking, since the new Friedmann equation contains the second derivative of the density. We believe it is necessary to provide a brief overview of the standard cosmological model in its analytical form (see Section IV). This is important because in Section V, we will present the numerical results and compare the new model with the standard one. In Section VI, we will draw our conclusions.

\section{The $f(R,\mathcal{L}_{m})$ theory}
\noindent
As a generalization of the standard Einstein-Hilbert Lagrangian in General Relativity, i.e.
\begin{equation}
\mathcal{L}_{GR}=\frac{1}{2\kappa}R+\mathcal{L}_{m},
\end{equation}
where $\kappa=8\pi G_{N}$, a new Lagrangian has been proposed by
making use of the invariants $R$ and $\mathcal{L}_{m}$.  It makes use
of a general function, $f(R,\mathcal{L}_{m})$, and reads simply
with an action \cite{FRLm,FRLm2,FRLm3}
\begin{equation}
S=\int d^{4}x \sqrt{-g} f(R,\mathcal{L}_{m}).
\end{equation}
From the formal variation of the action ($\delta_{g}S=0$) with respect
to the metric, the following field equations are obtained \cite{CosmologyFRLm,Harko2014}
\begin{equation}\label{FRLMFE}
f_{R}R_{\mu\nu} + (g_{\mu\nu}\Box-\nabla_{\mu}\nabla_{\nu})f_{R}-\frac{1}{2}(f-f_{L_{m}}\mathcal{L}_{m})g_{\mu\nu}=\frac{1}{2}f_{L_{m}}T_{\mu\nu},
\end{equation}
where $f_{R}=\frac{\partial f}{\partial R}$, $f_{L_{m}}=\frac{\partial
  f}{\partial L_{m}}$ and the energy-momentum tensor $T_{\mu\nu}$ is
given by
\begin{equation} \label{em}
T_{\mu\nu}=\frac{-2}{\sqrt{-g}}\frac{\delta(\sqrt{-g}\mathcal{L}_{m})}{\delta
  g^{\mu\nu}}.
\end{equation}
An alternative form of these field equations may be obtained by taking the trace of (\ref{FRLMFE}) and solving for $\Box f_{R}$, namely
\begin{equation} \label{second}
\Box f_{R}=\frac{1}{6}f_{L_{m}}T+\frac{2}{3}(f-f_{L_{m}}L_{m})-\frac{1}{3}f_{R}R.
\end{equation}
When replacing it back into (\ref{FRLMFE}), we obtain the equation
\begin{equation}
f_{R}G_{\mu\nu}+\frac{1}{6}(f+f_{R}R - f_{L_{m}}L_{m})g_{\mu\nu}-\frac{1}{2}f_{L_{m}}\left( T_{\mu\nu}-\frac{1}{3}T g_{\mu\nu}\right)-\nabla_{\mu}\nabla_{\nu}f_{R}=0.
\end{equation}
Here, $G_{\mu\nu}=R_{\mu\nu}-\frac{1}{2}g_{\mu\nu}R$ represents the Einstein
tensor. As mentioned in the Introduction, the particular form of
$f(R,\mathcal{L}_{m})$ we wish to work with is the same as the one presented in \cite{Lobato}
\begin{equation}
f(R,\mathcal{L}_{m})=\frac{R}{2\kappa}+\mathcal{L}_{m}+\sigma R\mathcal{L}_{m},
\end{equation}
where $\sigma$ is the parameter which determines the coupling between
matter and geometry.

With respect to the choice of the matter Lagrangian $\mathcal{L}_{m}$
that reproduces the perfect fluid Energy-Momentum tensor there is a
current debate \cite{EnergyMomentumHarko,  MinazzolliHarko}. While some authors make the
choice $\mathcal{L}_{m}=\pm\rho$, others suggest $\mathcal{L}_{m}=\pm P$ \cite{MatterLagrangian} with particular values for the polytropic index. With the choice in \cite{Lobato}, i.e. $\mathcal{L}_{m}=-P$, the field equations read
\begin{equation}\label{LobatoFE}
\left(\frac{1}{2\kappa}-\sigma P\right)R_{\mu\nu}-\sigma \left[g_{\mu\nu}\Box-\nabla_{\mu}\nabla_{\nu} \right]P-\frac{1}{4\kappa}R g_{\mu\nu}=\frac{1}{2}\left( 1+\sigma R\right)T_{\mu\nu},
\end{equation}
or in the alternative form
\begin{equation} \label{LobatoFE2}
(1-2\kappa\sigma P) G_{\mu\nu}+\frac{1}{3}Rg_{\mu\nu}-\frac{\kappa\sigma P}{3}Rg_{\mu\nu}-\kappa (1+\sigma R) \left(T_{\mu\nu}-\frac{1}{3}Tg_{\mu\nu} \right)+2\kappa\sigma\nabla_{\mu}\nabla_{\nu}P=0,
\end{equation}
which is consistent with the equations displayed above.  The conservation
of the energy-momentum tensor is given in general by \cite{Harko2014}
\begin{equation} \label{em2}
\nabla^{\mu}T_{\mu\nu}=2\nabla^{\mu}\ln \left[f_{L_{m}}(R,\mathcal{L}_{m}) \right]\frac{\partial \mathcal{L}_{m}}{\partial g^{\mu\nu}}.
\end{equation}
This suggests that choosing the matter Lagrangian to be proportional to
$\rho$ or $P$ leads to the conservation of the energy-momentum
tensor. However, this is only the case when relying directly on
equation (\ref{em}). Unfortunately, given the choices of the
matter Lagrangian as mentioned above, it does not reproduce the
perfect fluid energy-momentum tensor. To illustrate this point, if we
start with $\mathcal{L}_m=\pm \rho$, the energy-momentum tensor
defined in (\ref{em}) will not contain the pressure term.
On the other hand, by using a constrained variation in the metric \cite{ MatterLagrangian, MinazzolliHarko, Brown}  or alternatively, a
constrained Lagrangian formalism, we can explicitly show that, for example, choosing the energy density as the matter Lagrangian yields a perfect fluid energy-momentum tensor. However, its conservation no longer holds, as we introduce the metric through the constraints.

There is a certain confusion on this issue in the literature
\cite{Gonclaves}. This is a point of divergence between our approach and others. Therefore, we explicitly demonstrate the non-conservation of $T_{\mu \nu}$. The
equations of motion, as presented in
(\ref{FRLMFE}), (\ref{LobatoFE}), and (\ref{LobatoFE2}), remain valid when using the
constrained formalism. In this formalism, $T_{\mu \nu}$ is the
energy-momentum tensor of perfect fluid, which, in general, is not conserved and is defined as
\begin{equation} \label{em3}
T_{\mu\nu}=(\rho+P)u_{\mu}u_{\nu}-Pg_{\mu\nu}.
\end{equation}
As mentioned earlier, the model under discussion displays a
non-conservation of the energy-momentum tensor, specifically $\nabla_{\mu}T^{\mu\nu}\neq 0$ (at least in general). This
non-conservation is also a characteristic of other gravity models such as Rastall Gravity \cite{Rastall,Rastall1}, and has been advocated in modular gravity models \cite{Sudarsky,Sudarsky1}.
From the field equations (\ref{FRLMFE}) we can
obtain the modified Friedmann equations in the flat ($k=0$)
Friedmann--Lemaitre--Robertson--Walker metric
\begin{equation}
ds^{2}=g_{\mu\nu}dx^{\mu}dx^{\nu}=dt^{2}-a^{2}(t)\left(dr^{2}+r^{2}d\theta^{2}+r^{2}\sin^{2}\theta d\phi^{2} \right).
\end{equation}
By taking the 0-0 components of the new field equations, one obtains \cite{CosmologyFRLm}
\begin{equation}
3H^{2}f_{R}+\frac{1}{2}(f-f_{R}R-f_{L_{m}}\mathcal{L}_{m})+3H\dot{f}_{R}=\frac{1}{2}f_{L_{m}}\rho,
\end{equation}
where $H$ is the Hubble parameter given by $H=\dot{a}/a$.
The space components (i.e. $\mu=i$, $\nu=j$) give
\begin{equation}
\dot{H}f_{R}+3H^{2}f_{R}-\ddot{f}_{R}-3H\dot{f}_{R}+\frac{1}{2}(f_{L_{m}}\mathcal{L}_{m}-f)=\frac{1}{2}f_{L_{m}}P,
\end{equation}
where the $\ddot{f}_{R}$ term comes from the double derivatives in the field equations ($\Box$ and $\nabla_{\mu}\nabla_{\nu}$). For the model we have chosen, we have
\begin{equation}
f_{R}=\frac{1}{2\kappa}+\sigma \mathcal{L}_{m}, \qquad f_{L_{m}}=1+\sigma R, \qquad \dot{f}_{R}=\sigma \dot{\rho}, \qquad \ddot{f}_{R}=\sigma \ddot{\rho}.
\end{equation}
Together with the choice $\mathcal{L}_{m}=\rho$, the new Friedmann equations read
\begin{eqnarray}
H^{2}&=&\kappa\frac{\rho}{3}+\sigma\kappa\left[ 4\dot{H}\rho+6H^{2}\rho-2H\dot{\rho}\right],\label{F1FRL}\\
\frac{\ddot{a}}{a}&=&-\frac{\kappa}{6}\left(\rho + 3P \right)-\sigma\kappa [ \dot{H}(3P+\rho)+6H^{2}P +\ddot{\rho}+2H\dot{\rho}].\label{F2FRL}
\end{eqnarray}
In general relativity, the two Friedmann equations lead to the
conservation law $\dot{\rho}+3H(\rho+P)= 0$. It is known that the
same conservation law follows from $\nabla^{\mu}T_{\mu \nu}=0$
\cite{Einstein}.  It is straightforward to derive from (\ref{F1FRL})
and (\ref{F2FRL}) the following relation
\begin{equation}\label{ContFRL}
\dot{\rho}+3H(\rho+P)=-6\sigma\left[H\dot{H}(3P+11\rho)+\dot{\rho}(3H^{2}+\dot{H})+6H^{3}(P+\rho) +2\ddot{H}\rho\right],
\end{equation}
which clearly gives back the conservation law (and therefore, the
conservation of the energy-momentum tensor) in the case
$\sigma=0$. The nature of the non-conservation of the energy-momentum
tensor is explicitly proved here. It should have been addressed also in
\cite{AcceleratingFRLm, ViscousDM, Gonclaves} where the authors
consider cosmological models of $f(R, L_m)$ theories.

Commonly, equations (\ref{F1FRL}) and (\ref{F2FRL}) are known as (modified)
Friedmann equations. In some papers \cite{Singh} equation (\ref{F2FRL}) is also
called Raychaudhuri equation. Both equations are related by (\ref{ContFRL}).

In rewriting the Friedmann equations in a dimensionless form, specific choices were made to align the theoretical framework with observational benchmarks and simplify the numerical analysis. By introducing \(\chi = \frac{\rho}{\rho_{\rm crit}}\), we scale the density to the critical density, which is pivotal in determining the curvature of the universe. Similarly, scaling the Hubble parameter by its current value \(H_0\) through \(h^2 = \frac{H^2}{H_0^2}\) allows us to examine the evolution of the expansion rate relative to its present value. The introduction of \(\tilde{\sigma} = H_0^2 \sigma\) helps in assessing the impact of the coupling constant \(\sigma\) within the typical energy scales of the current universe. Lastly, by using \(\eta = H_0 t\), time derivatives are normalized to the current expansion rate, making them dimensionless and easier to handle numerically. These transformations are not only mathematically convenient but also provide a direct link between theoretical predictions and observable quantities. To this end, we introduce
\begin{equation}
\chi \equiv \frac{\rho}{\rho_{\rm crit}} =\frac{\kappa
  \rho}{3H_{0}^{2}}, \qquad h^{2}\equiv \frac{H^{2}}{H_{0}^{2}},
\qquad \tilde{\sigma}\equiv H_{0}^{2}\sigma,\qquad \eta=H_{0} t,
\qquad \rho_{crit}=3H_0^2/\kappa,
\end{equation}
where $\rho_{0}$ is a reference value of $\rho$ at the present epoch.
Similar comment applies to the Hubble parameter $H_0$. In addition, we
write the time derivatives in terms of derivatives with respect to the
dimensionless parameter $\eta$ by means of
\begin{equation}
\accentset{\circ}{A}\equiv \frac{dA}{d\eta}=\frac{dA}{dt}\frac{dt}{d\eta}=\dot{A}\frac{1}{H_{0}}.
\end{equation}

In the context of our model, the barotropic index \(\gamma\)
characterizes the equation of state (EOS) of the cosmological fluid,
namely \(P = (\gamma - 1) \rho\). This parameter is crucial for
defining the thermodynamic properties of the universe's contents,
where \(\gamma = 1\) corresponds to a universe dominated by
non-relativistic matter (the choice we make here for the calculations), \(\gamma = 4/3\) to one dominated by radiation, and \(\gamma = 0\) to the cosmological constant scenario with vacuum energy. Each value of \(\gamma\) distinctly affects the evolution dynamics of the universe, as reflected in the modified Friedmann equations. By means of the aforementioned (EOS), equations
(\ref{F1FRL}), (\ref{F2FRL}) and (\ref{ContFRL}) take the form
\begin{eqnarray}
h^{2}&=&\chi + \tilde{\sigma}[ 18h^{2}\chi + 12\accentset{\circ}{h}\chi-6h\accentset{\circ}{\chi}],\label{F1DL}\\
h^{2}+\accentset{\circ}{h}&=&-\frac{1}{2}(3\gamma-2)\chi-\tilde{\sigma}[3(3\gamma-2)\accentset{\circ}{h}\chi+18(\gamma-1)h^{2}\chi+3\accentset{\circ \circ}{\chi}+6h\accentset{\circ}{\chi}],\label{F2DL}\\
3\accentset{\circ}{\chi}+9\gamma h\chi&=& -6\tilde{\sigma}[(9\gamma+24)h\accentset{\circ}{h}\chi+3\accentset{\circ}{\chi}(3h^{2}+\accentset{\circ}{h})+18\gamma h^{3}\chi+6\accentset{\circ \circ}{h}\chi].\label{ContDL}
\end{eqnarray}
Since we want to compare the modified cosmological model with standard
cosmology  we should also include the cosmological constant
$\Lambda$. This can be done without much effort by considering the action
\begin{equation}
S_{f(R,\mathcal{L}_{m})+\Lambda}=\int d^{4}x \sqrt{-g} \left(f_{\Lambda}(R,\mathcal{L}_{m})-\frac{\Lambda}{\kappa} \right)
\end{equation}
with
\begin{equation}
f_{\Lambda}(R,\mathcal{L}_{m})=\frac{R}{2\kappa}+\mathcal{L}_{m}+\sigma R\mathcal{L}_{m}.
\end{equation}
This gives the new Einstein equations
\begin{equation}
f_{R}R_{\mu\nu}+(g_{\mu\nu}\Box -\nabla_{\mu}\nabla_{\nu})f_{R}-\frac{1}{2}(f-f_{L_{m}}L_{m})g_{\mu\nu}+\frac{\Lambda}{2\kappa}g_{\mu\nu}=\frac{1}{2}f_{L_{m}}T_{\mu\nu}
\end{equation}
as well as the generalized Friedmann equations
\begin{eqnarray}
3H^{2}f_{R}+\frac{1}{2}(f-f_{L_{m}}\mathcal{L}_{m}-f_{R}R)+3H\dot{f}_{R}-\frac{\Lambda}{2\kappa}&=&\frac{1}{2}f_{L_{m}}\rho,\\
\dot{H}f_{R}+3H^{2}f_{R}-\ddot{f}_{R}-3H\dot{f}_{R}+\frac{1}{2}(f_{L_{m}}\mathcal{L}_{m}-f)+\frac{\Lambda}{2\kappa}&=&\frac{1}{2}f_{L_{m}}P.
\end{eqnarray}
With the explicit choice of $f_{\Lambda}$ the modified Friedmann equations
simplify to
\begin{eqnarray}
H^{2}&=&\kappa\frac{\rho}{3}+\sigma\kappa\left[ 4\dot{H}\rho+6H^{2}\rho-2H\dot{\rho}\right] + \frac{\Lambda}{3},\label{F1FRLambda}\\
\frac{\ddot{a}}{a}&=&-\frac{\kappa}{6}\left(\rho + 3P \right)-\sigma\kappa [ \dot{H}(3P+\rho)+6H^{2}P +\ddot{\rho}+2H\dot{\rho}]+\frac{\Lambda}{3}.\label{F2FRLambda}
\end{eqnarray}
Their dimensionless form can be written by using the dimensionless parameter $\Omega_{\Lambda}=\rho_{\rm vac}/\rho_{\rm crit}$ as follows
\begin{eqnarray}
h^{2}&=&\chi + \tilde{\sigma}[ 18h^{2}\chi + 12\accentset{\circ}{h}\chi-6h\accentset{\circ}{\chi}]+\Omega_{\Lambda},\label{F1DLnew}\\
h^{2}+\accentset{\circ}{h}&=&-\frac{1}{2}(3\gamma-2)\chi-\tilde{\sigma}[3(3\gamma-2)\accentset{\circ}{h}\chi+18(\gamma-1)h^{2}\chi+3\accentset{\circ \circ}{\chi}+6h\accentset{\circ}{\chi}]+\Omega_{\Lambda}.\label{F2DLnew}
\end{eqnarray}
These equations are now suitable for numerical integration.

Normally, the inclusion of $\Lambda$ in a modified gravity theory
might not be well justified as the modification itself can
account
for Dark Energy. But it appears that this is a too global statement
and the role of $\Lambda$ can vary from model to model. In the
section on numerical results we will demonstrate that the cosmological
constant does not affect the characteristic feature of the model,
i.e., its shortcoming to produce a reasonable lifetime of the universe.

\section{The initial values and the $\sum_i \Omega_i=1$ relation}
\noindent
The Friedman equations are now first order in $h$ and second order in
$\chi$. Therefore, we need three initial values, which we provide at the time $t_{0}$ corresponding to the present epoch. We have
\begin{equation}
h_{0}\equiv h(t_{0})=\frac{H_{0}}{H_{0}}=1, \quad \chi_{0}\equiv\chi (t_{0})=\frac{\kappa \rho_{0}}{3H_{0}^{2}}, \quad \accentset{\circ}{h}_{0}\equiv\accentset{\circ}{h}(t_{0})=\frac{\dot{H}_{0}}{H^{2}_{0}}, \quad \accentset{\circ}{\chi}_{0}\equiv\accentset{\circ}{\chi}(t_{0})=\frac{\kappa \dot{\rho}_{0}}{3 H_{0}^{3}}.
\end{equation}
We look for model independent measurements for the density of the Universe ($\rho_{0}$), the Hubble parameter ($H_{0}$) and its first and second derivatives ($\dot{H}_{0}$, $\ddot{H}_{0}$). For $\dot{H}$ we use the deceleration parameter $q$, given by \cite{JerkSnap}
\begin{equation}
q\equiv-\frac{\ddot{a}a}{\dot{a}^{2}}, \quad \rightarrow \quad\frac{\dot{H}}{H^{2}}=-(1+q)
\end{equation}
and thus,
\begin{equation} \label{Hdot}
\dot{H}_{0}=-H_{0}^{2}(1+q_{0}),
\end{equation}
where $q_{0}=q(t=t_{0})$ has been obtained in a model independent way
\cite{Cosmographic}. The jerk parameter will be useful for
$\ddot{H}_{0}$ appearing in the second modified Friedmann equation. This is given by \cite{JerkSnap,VisserM}
\begin{equation}
j=\frac{\dddot{a}}{aH^{3}} \rightarrow j=\frac{\ddot{H}}{H^{3}}+3\frac{\dot{H}}{H^{2}}+1
\end{equation}
and therefore,
\begin{equation}
\ddot{H}_{0}=(j_{0}-1)H_{0}^{3}-3\dot{H}_{0}H_{0}=(j_{0}+q_{0}+2)H_{0}^{3}.
\end{equation}
To ensure that our exploration of the modified \(f(R, \mathcal{L}_m)\) gravity model remains grounded in observational reality, we adopt standard cosmological parameters that are widely accepted within the community. This choice allows us to rigorously test whether the modifications introduced by the theoretical framework can provide a viable alternative to the \(\Lambda\)CDM model, particularly in light of recent tensions and discrepancies such as those in measurements of the Hubble constant. By using these parameters, the model's predictions can be directly compared with those derived from both local and cosmological scales, providing a comprehensive evaluation of its empirical adequacy. For the numerical values, it is important to note that due to recent
discrepancies in measurements of the Hubble constant, known as the
Hubble Tension \cite{HubbleTension}, there are several values for $H_{0}$. For example we have
$H^{+}_{0}=$ 73,52$\pm$1,62 km/s/Mpc and $H^{-}_{0}=$ 67,4 $\pm$0,5
km/s/Mpc, respectively. As for the current energy density
of the Universe and the deceleration parameter, we use values found in
\cite{Cosmographic,TransitFRLm,AcceleratingFRLm}:
$\chi_{0}=\frac{\rho_{0}}{\rho_{\rm crit}}=0.285\pm0.012$, $q_{0}$ =
-0.545 $\pm$ 0.107, $j_{0}=1.30\pm 0.37$ with the value for
the Hubble parameter taken as $H_{0}=71.34\pm1.74$ km s$^{-1}$
Mpc$^{-1}$. It is worth noting that these are specific parameter
choices, but other sets of parameters are available in
\cite{Cosmographic}. Possible values for these parameters are
displayed in Table~\ref{Table1}. This particular choice of parameters that as of today we are as close as possible to the standard cosmology.
\squeezetable
\begin{table}[]
\begin{tabular}{|lllll|}
\hline
\multicolumn{5}{|c|}{Density Parameters}                                                                                               \\ \hline
\multicolumn{1}{|l|}{Data}  & \multicolumn{2}{l|}{SNIa}                           & \multicolumn{2}{l|}{Hubble}                        \\ \hline
\multicolumn{1}{|l|}{$\Omega_{m,0}$}      & \multicolumn{2}{l|}{0.285$\pm$0.012}                               & \multicolumn{2}{l|}{0.239$\pm$0.015}                              \\ \hline
\multicolumn{1}{|l|}{$\Omega_{\Lambda,0}$}      & \multicolumn{2}{l|}{0.715}                               & \multicolumn{2}{l|}{0.761}                              \\ \hline
\multicolumn{1}{|l|}{$\Delta$}      & \multicolumn{2}{l|}{0.012}                               & \multicolumn{2}{l|}{0.015}                              \\ \hline
\multicolumn{5}{|c|}{Hubble data}                                                                                                      \\ \hline
\multicolumn{1}{|c|}{Model} & \multicolumn{1}{c|}{$H_{exp}$} & \multicolumn{1}{c|}{GP} & \multicolumn{1}{c|}{GA} & \multicolumn{1}{c|}{$\Lambda$CDM} \\ \hline
\multicolumn{1}{|l|}{$H_{0}$}     & \multicolumn{1}{l|}{73.88$\pm$1.34}     & \multicolumn{1}{l|}{73.44$\pm$1.40}   & \multicolumn{1}{l|}{71.34$\pm$1.74}   &    72.08$\pm$1.06                      \\ \hline
\multicolumn{1}{|l|}{$q_{0}$}     & \multicolumn{1}{l|}{-1.070$\pm$0.093}     & \multicolumn{1}{l|}{-0.856$\pm$0.111}   & \multicolumn{1}{l|}{-0.545$\pm$0.107}   & -0.645$\pm$0.023                         \\ \hline
\multicolumn{1}{|l|}{$j_{0}$}     & \multicolumn{1}{l|}{3.00$\pm$0.62}     & \multicolumn{1}{l|}{1.30$\pm$0.37}   & \multicolumn{1}{l|}{0.52$\pm$0.24}   &              1.00            \\ \hline
\multicolumn{5}{|c|}{Pantheon data}                                                                                                    \\ \hline
\multicolumn{1}{|c|}{Model} & \multicolumn{1}{c|}{$H_{exp}$} & \multicolumn{1}{c|}{GP} & \multicolumn{1}{c|}{GA} & \multicolumn{1}{c|}{$\Lambda$CDM} \\ \hline
\multicolumn{1}{|l|}{$H_{0}$}     & \multicolumn{1}{l|}{71.13$\pm$0.46}     & \multicolumn{1}{l|}{71.92$\pm$0.38}   & \multicolumn{1}{l|}{71.81$\pm$1.14}   &         71.84$\pm$0.22                 \\ \hline
\multicolumn{1}{|l|}{$q_{0}$}     & \multicolumn{1}{l|}{-0.616$\pm$0.105}     & \multicolumn{1}{l|}{-0.558$\pm$0.040}   & \multicolumn{1}{l|}{-0.466$\pm$0.244}   &         -0.572$\pm$0.018                 \\ \hline
\multicolumn{1}{|l|}{$j_{0}$}     & \multicolumn{1}{l|}{1.56$\pm$0.74}     & \multicolumn{1}{l|}{0.85$\pm$0.12}   & \multicolumn{1}{l|}{0.55$\pm$1.65}   &             1.00             \\ \hline
\end{tabular}
\caption{Specific choices of the parameters according to \cite{HubbleTension,Cosmographic,TransitFRLm,AcceleratingFRLm}.}
\label{Table1}
\end{table}
For a standard reference value of $\accentset{\circ}{\chi}_{0}$, we consider the standard cosmological model and use the continuity equation (\ref{ContDL}) with $\tilde{\sigma}=0$, even in the presence of $\Lambda\neq0$. This leads to
\begin{equation}
3\accentset{\circ}{\chi}+9\gamma h \chi=0.
\end{equation}
For the initial value we then obtain
\begin{equation}
\accentset{\circ}{\chi}_{0}=-3\gamma h_{0}\chi_{0}=-3\gamma \chi_{0}.
\end{equation}
Since $h_{0}=1$, in the case of dust $\gamma=1$, we have
\begin{equation} \label{initialrhodot}
\accentset{\circ}{\chi}_{0}=-3\chi_{0}.
\end{equation}
In mathematics, when solving differential equations, one typically
assumes that all parameters and functions entering the equation, as
well as the initial values are
known and given. However, in physics, the situation can become more
complex because these parameters and initial values may not be
precisely known. When considering a differential equation at the point
where the initial values are applied, it leads to a relationship
between
these initial values and
the parameters of the equation. The Friedmann equations provide a notable example of this phenomenon.

The first modified Friedmann equation can be expressed in terms of the density
parameters $\Omega_i$. To this purpose, let us start by defining the $\Omega_{m}$ parameter as
\begin{equation}
\Omega_{m}=\frac{\rho}{\rho_{\rm crit}}=\frac{\kappa \rho}{3H^{2}}.
\end{equation}
This, in turn, implies that $\chi=\Omega_{m}$ and $\rho=\Omega_{m} \rho_{\rm crit}=\Omega\frac{3H^{2}}{\kappa}$. Thus, our first Friedmann equation can be reformulates as follows
\begin{equation}
H^{2}=\Omega_{m} H^{2}+\Omega_{m} H^{2}(18\sigma H^{2})+\Omega_{m}
H^{2} (12\sigma \dot{H})-\sigma \kappa H \dot{\rho} +\frac{\Lambda}{3}.
\end{equation}
Recognizing this equation as
\begin{equation}
\Omega_{m} + \Omega_{\Lambda}+ \tilde{\sigma} \left[18 \Omega_{m} H^{2}+12 \Omega_{m} \dot{H}-\frac{\kappa \dot{\rho}}{H} \right]=1,
\end{equation}
we can express it in a more compact form. Finally, defining the term proportional to $\sigma$ as $\Omega_{\sigma}$, we arrive at
\begin{equation}\label{OmegaSigma}
\Omega_{m,0} +\Omega_{\Lambda, 0}+\Omega_{\sigma,0}=1,
\end{equation}
where we have taken the values at $t=t_0$. Making use of (\ref{Hdot}) and
 (\ref{initialrhodot}) the above equation reduces to
\begin{equation} \label{Omega}
\Omega_{m,0} + \Omega_{\Lambda,0}+\tilde{\sigma}\Omega_{m,0} (15-12 q_0)=1.
\end{equation}
Equation (\ref{Omega}) establishes a relationship between initial
values and parameters within the modified Friedmann equation. Moreover, it
introduces the deceleration parameter $q_0$, which is a measured
quantity,
though not strictly
necessary for solving the first order modified Friedmann equation in $H$.
Consequently, there are generally two approaches to solve the new
Friedmann equations.  The first one involves specifying $h_0=1$, $\chi_0$,
$\accentset{\circ}{\chi}_{0}$, and $\Omega_{\Lambda}$ while varying $\tilde{\sigma}$. Equation (\ref{Omega}) can then be used to determine $q_0$. The second approach relies on the measured value of $q_0$ and fixes the other parameters
as previously described. In this case, equation (\ref{Omega}) provides
the parameter $\tilde{\sigma}$. In particular, we note that in the case without coupling between $R$ and $\mathcal{L}_{m}$, the relative densities today read
\begin{equation}
\Omega_{m,0}+\Omega_{\Lambda,0}=1.
\end{equation}

Numerically, however, there is an uncertainty in the obtained values,
reflected in the relation $\Omega_{m,0}+\Omega_{\Lambda,0}=1\pm \Delta$, where $\Delta$ represents this uncertainty. We can exploit it in our model by setting, as shown in equation (\ref{OmegaSigma}),
\begin{equation} \textcolor{red}{\label{uncertainty_new}}
\tilde{\sigma}=\pm \frac{\Delta}{\Omega_{m,0}(15-12q_{0})}.
\end{equation}

\section{The standard cosmology}
Before embarking on a discussion of the numerical solutions it is
convenient to briefly outline the analytical picture of the standard
cosmology. This makes sense since we will be comparing the two scenarios.

The dimensionless solution for $h_{\Lambda+}$ for the standard
cosmological model can be obtained by taking $\tilde{\sigma}=0$
in equations (\ref{F1DL}) and (\ref{F2DL}) with $\gamma=1$, namely
\begin{eqnarray}
h^{2}&=&\chi +\Omega_{\Lambda},\\
h^{2}+\accentset{\circ}{h}&=&-\frac{1}{2}\chi +\Omega_{\Lambda}.
\end{eqnarray}
Combining the above equations, we arrive at a Riccati equation of the form
\begin{equation}
\accentset{\circ}{h}=-\frac{3}{2} h^{2} + \frac{3}{2}\Omega_{\Lambda}.
\end{equation}
We can solve the differential equation above by setting
\begin{equation}
\int_{h'=1}^{h'=h} \frac{dh'}{\Omega_{\Lambda}-(h')^{2}}=\frac{3}{2}\int_{\eta'= \eta_{0}}^{\eta'=\eta} d\eta',
\end{equation}
where $\eta_{0}=H_{0}t_{0}$ is the time at which the
Hubble parameter is $H_{0}$, which, in turn, is related to the present values of the parameters used. To solve the integral above, we redefine $x=\frac{h'}{\sqrt{\Omega_{\Lambda}}}$ and write
\begin{equation}\label{IntStan}
\frac{1}{\sqrt{\Omega_{\Lambda}}}\int_{x=1/\sqrt{\Omega_{\Lambda}}}^{x=h/\sqrt{\Omega_{\Lambda}}} \frac{dx}{1-x^{2}}=\frac{3}{2}(\eta-\eta_{0}).
\end{equation}
One possible solution for the integral on the left-hand side is
\begin{equation}
\int \frac{dx}{1-x^{2}}=\frac{1}{2}\ln \left(\frac{1+x}{1-x}\right).
\end{equation}
However, such a solution is valid for $-1<x<1$. This would imply that $h<\sqrt{\Omega_{\Lambda}}$ and thus, $\chi=h^{2}-\Omega_{\Lambda}<0$, which would suggest an unphysical negative density. Another solution is
\begin{equation}
\int \frac{dx}{1-x^{2}}=\frac{1}{2}\ln \left(\frac{x+1}{x-1} \right).
\end{equation}
It is valid whenever $|x|>1$. Moreover, it implies $h^{2}>\Omega_{\Lambda}$ and, consequently $\chi=h^{2}-\Omega_{\Lambda}>0$, ensuring a positive and physically meaningful density. If we also use the fact that
\begin{equation}
\coth^{-1}x=\frac{1}{2}\ln \left(\frac{x+1}{x-1} \right)
\end{equation}
and solve for the integral (\ref{IntStan}), we end up with the result
\begin{equation}
\frac{1}{\sqrt{\Omega_{\Lambda}}}\left[\coth^{-1}\left(\frac{h}{\sqrt{\Omega_{\Lambda}}} \right)-\coth^{-1}\left(\frac{1}{\sqrt{\Omega_{\Lambda}}} \right) \right]=\frac{3}{2}(\eta-\eta_{0}).
\end{equation}
Solving for $h$ in the $\gamma=1$ case leads to
\begin{equation}
h_{\Lambda+}(\eta)=\sqrt{\Omega_{\Lambda}}\coth \left[\frac{3}{2}\sqrt{\Omega_{\Lambda}}(\eta-\eta_{0})+\coth^{-1}\left(\frac{1}{\sqrt{\Omega_{\Lambda}}} \right) \right].
\end{equation}
For the general $\gamma$ case, the equation above reads
\begin{equation}
h_{\Lambda+}(\eta)=\sqrt{\Omega_{\Lambda}}\coth\left[\frac{3\gamma}{2}\sqrt{\Omega_{\Lambda}}(\eta-\eta_{0})+\coth^{-1}\left(\frac{1}{\sqrt{\Omega_{\Lambda}}}\right) \right].
\end{equation}
In the standard notation of the scale factor $a_{\Lambda}(t)$ and the Hubble parameter $H_{\Lambda}(t)$, we get
\begin{eqnarray}
a_{\Lambda+}(t)&=&\left(\frac{1-\Omega_{\Lambda}}{\Omega_{\Lambda}}\right)^{\frac{1}{3\gamma}}
\sinh^{\frac{2}{3\gamma}}{(\mu t+\nu)},\label{solutiona2}\\
H_{\Lambda+}(t)&=& \sqrt{\frac{\Lambda}{3}}
\coth{(\mu t+\nu)} \label{solutionH2},
\end{eqnarray}
where
\begin{equation}
\mu=\frac{\gamma\sqrt{3\Lambda}}{2},\quad
\nu=\frac{1}{2}\ln\left(\frac{1+\sqrt{\Omega_{\Lambda}}}{1-\sqrt{\Omega_{\Lambda}}}\right)-\mu t_0
\end{equation}
for a positive cosmological constant $\Lambda$. For completeness, we
also give the solution for $\Lambda<0$ which reads
\begin{eqnarray}
a(t)_{\Lambda-}&=&a_0\left[\frac{\cos{(\alpha-\beta t)}}{\cos{(\alpha-\beta t_0)}}\right]^\frac{2}{3\gamma},\\
H_{\Lambda-}(t)&=&\sqrt{-\frac{\Lambda}{3}}\tan{(\alpha-\beta t)}
\end{eqnarray}
with
\begin{equation} \label{alpha}
\alpha=\beta t_0+\tan^{-1}{\left(\sqrt{-\frac{3}{\Lambda}}H_0\right)},\quad\beta=\frac{\gamma\sqrt{-3\Lambda}}{2}.
\end{equation}

Without loss of generality we can set $t_0=0$. Then from (\ref{alpha})
we can infer that $0 <\alpha < \pi/2 $ and hence $\cos(\alpha)>0$. To
ensure that the scale factor is positive it suffices to choose $t \in
[\frac{\alpha -\pi/2}{\beta}, \frac{\alpha +\pi/2}{\beta}]$
  corresponding to Big Bang and collapse.

We will explore later how this simple solutions evolves when we introduce a non-zero $\sigma$.

Two important model-independent restrictions on a realistic
cosmological model {relate to the universe lifetime. In the
standard cosmological model, we can calculate the lifetime (neglecting
the short radiation period) by setting the argument of the Hubble
function to zero, which results in a singularity. Choosing $\eta_0=0$,
we obtain
\begin{equation} \label{lifetime}
  |\eta_{univ}|=
  \frac{1}{3\sqrt{\Omega_{\Lambda}}}\ln \left(\frac{1+\sqrt{\Omega_{\Lambda}}
    }{1- \sqrt{\Omega_{\Lambda}}}\right),
\end{equation}
we get $|\eta_{univ}| \simeq 0.96$ for $\Omega \simeq 0.7$. This
lifetime is not in conflict with model-independent estimates,
such as those from the oldest galaxies \cite{galaxies}
(which indicate the existence of galaxies some 400 million years after
the Big Bang) and \cite{uranium},
which points to a lower limit of 12.5Gy.

\section{Numerical results}
Assuming $\sigma \neq 0$, we have converted the system under discussion
into a first order system which we solved by standard numerical
methods. As an independent check, we used MAPLE18 routines and found that the
results coincide.

When solving the modified Friedmann equations numerically, the question arises:
What values should we take for $\tilde{\sigma}$? The standard
cosmology has two constants, the Newtonian one $G_N$ and the
cosmological constant. Interestingly, the dimension of $\sigma$ is the same as that of the Newtonian constant.  If we take $\sigma$ as
$nG_N$ with $n$ some number of order $1$, $\tilde{\sigma}$ would turn out extremely small. Therefore, it makes sense to initially} try more moderate values for $\tilde{\sigma}$. In the subsequent figures, the blue curve represents the standard cosmological model prediction.

\begin{figure}[ht]
\centering
\includegraphics[scale=0.5]{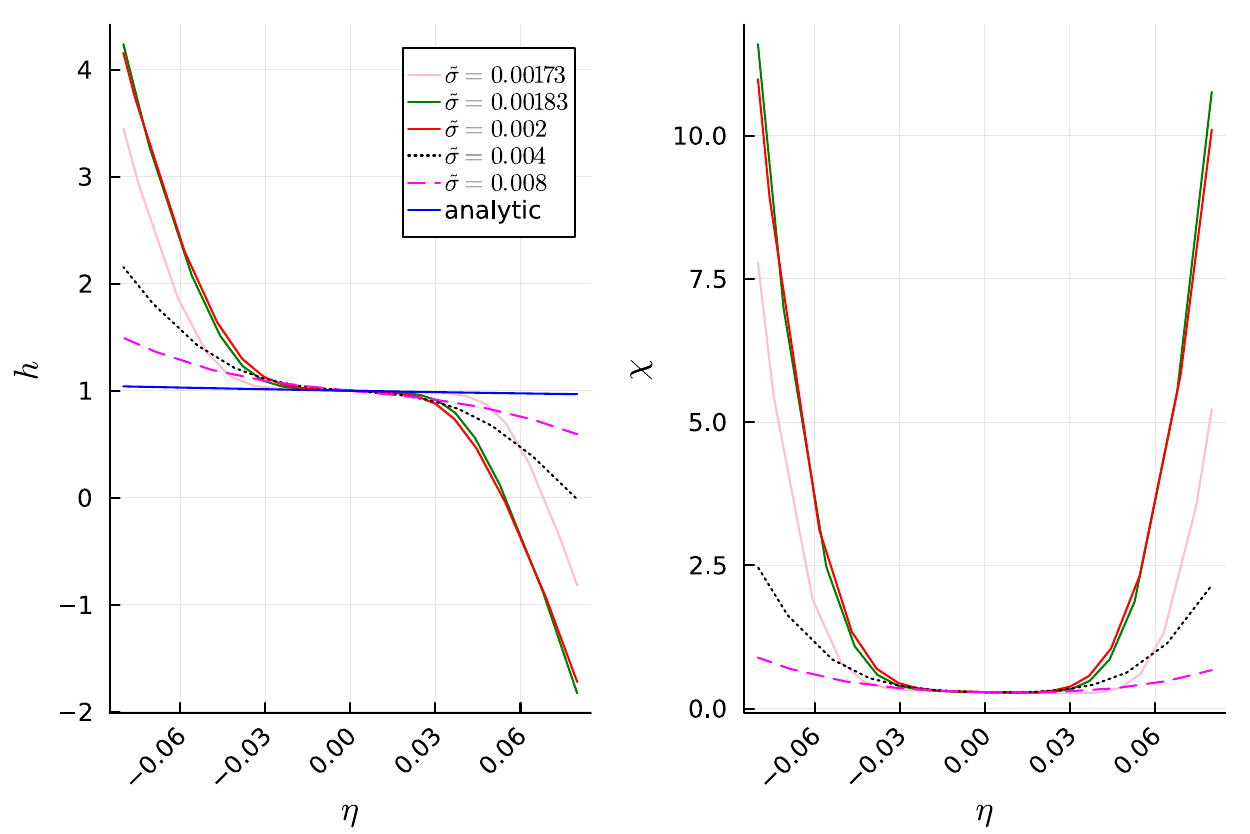}
\caption{The dimensionless Hubble function $h$ and the dimensionless
  density $\chi$ versus the dimensionless time $\eta$ for a choice of
  the parameter $\tilde{\sigma}$.}\label{span3}
\end{figure}
In Figure~\ref{span3}, we have chosen the initial values as discussed
in Section III, with $\Omega_{\Lambda}=0.7$. The chosen values of
$\tilde{\sigma}$ are relatively small to illustrate that the
dimensionless Hubble function closely follows the one of the standard
cosmological model around $\eta$ close to zero.
Several notable features can be observed in the plot, which are likely to persist for other values of $\tilde{\sigma}$. First of all, the Hubble function reaches the singularity much earlier than in the standard model. This poses a challenge if we aim to adapt the model under discussion as a viable cosmological model of our universe. As mentioned earlier, one of the constraints on any cosmological model is its ability to be roughly compatible with a lower limit on the lifetime of the universe. This constraint arises from evidence such as the existence of galaxies just 400 million years after the Big Bang \cite{galaxies}. Therefore, any cosmological model must strive to be in agreement with the estimated lifetime of the universe as predicted by the standard cosmological model.
Secondly, we notice another interesting feature. The Hubble function becomes negative which implies a contraction of the universe. Therefore, the dimensionless density $\chi$ starts increasing
again. This is known as the Big Crunch scenario which happens also
in $f(R, L_m)$ models with the simple choice $L_m=\rho$. The Big Crunch here is a result of the model and cannot be avoided.

\begin{figure}[ht]
\centering
\includegraphics[scale=0.5]{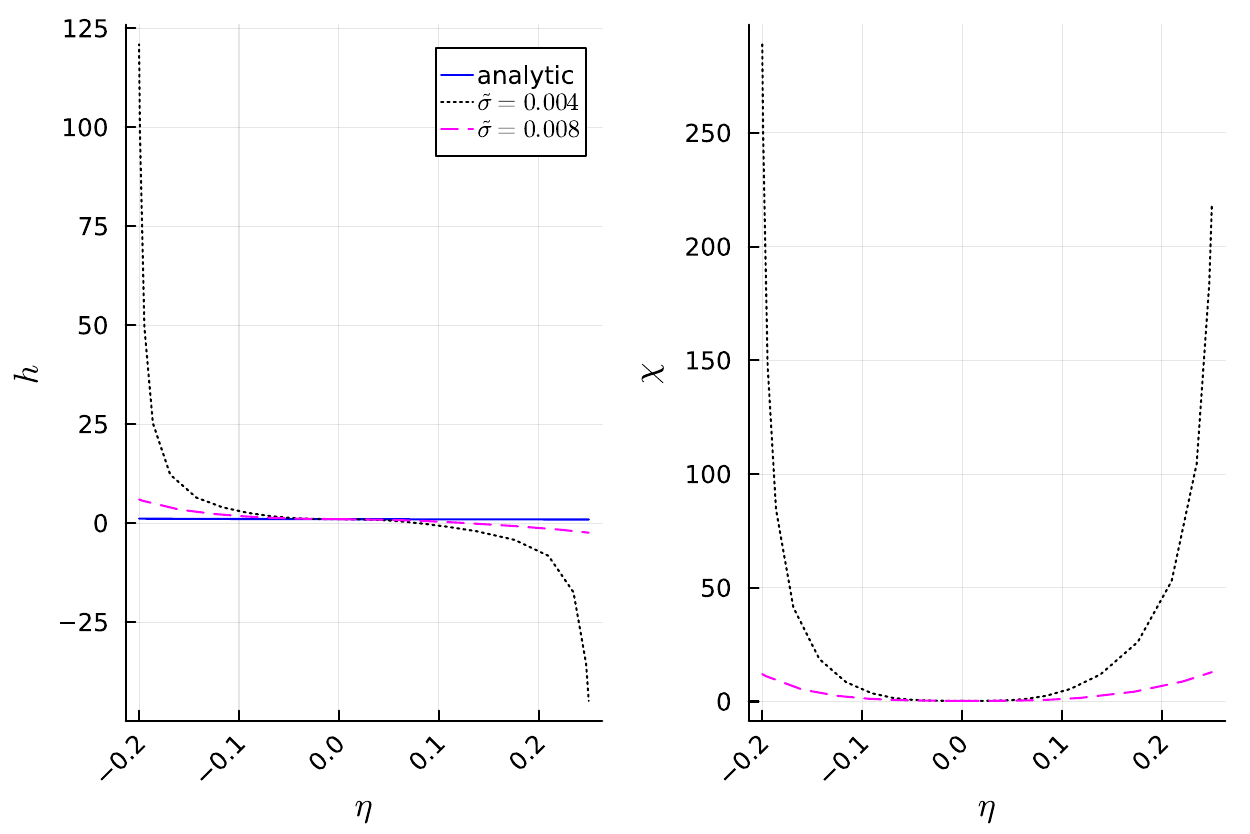}
\caption{The same plot as in \ref{span3}, but over a large timescale.}\label{span8}
\end{figure}

In Figure~\ref{span8}, we have plotted the same functions for two
values of $\tilde{\sigma}$ appearing also in figure
\ref{span3}, but over a larger scale of time. One can appreciate the
steep rise of $h$ at a relatively modest negative values of $\eta$.

\begin{figure}[ht]
\centering
\includegraphics[scale=0.5]{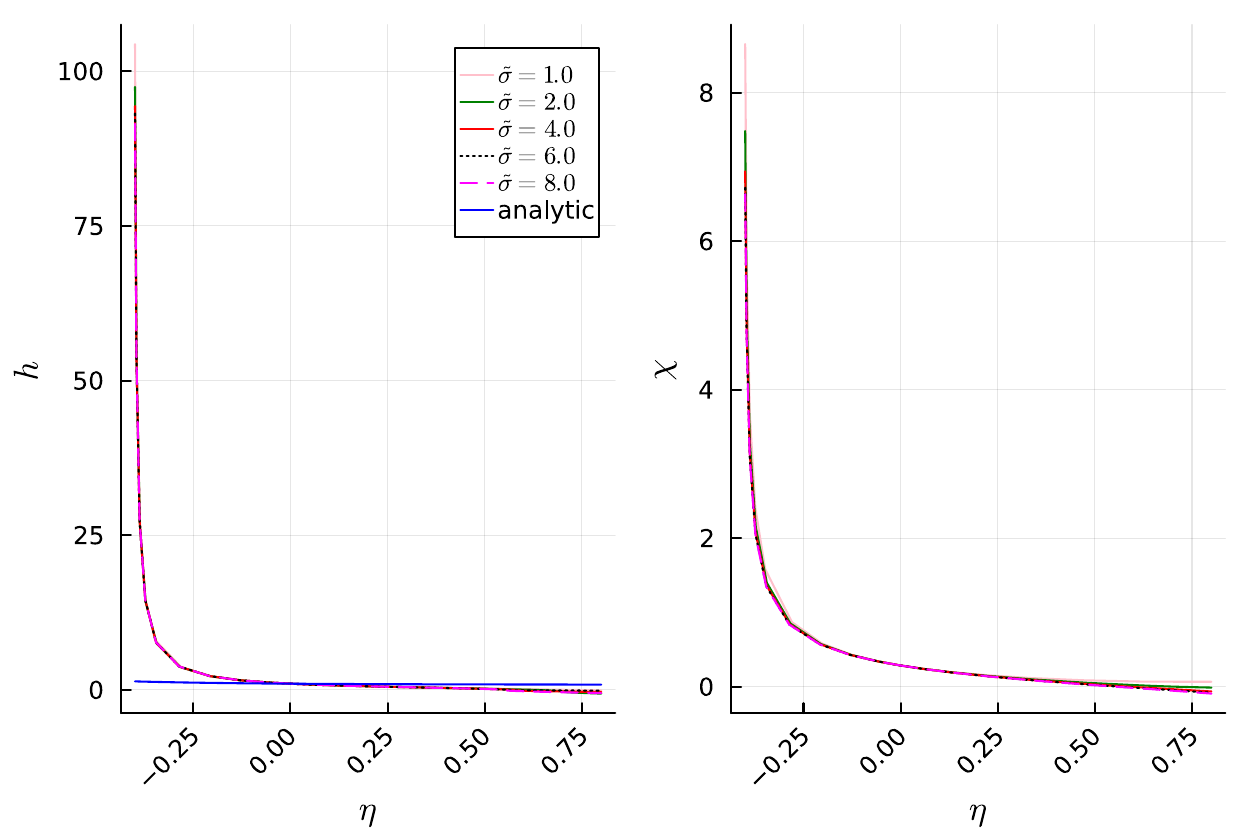}
\caption{The same plot as in \ref{span3}, but for large values of
  $\tilde{\sigma}.$.}\label{span6}
\end{figure}

In Figure~\ref{span6}, we have increased the values of
$\tilde{\sigma}$ but there is  no visible effect on the time of the
Big-Bang. However, for larger values of the parameter, the density becomes unphysically negative.

\begin{figure}[ht]
\centering
\includegraphics[scale=0.5]{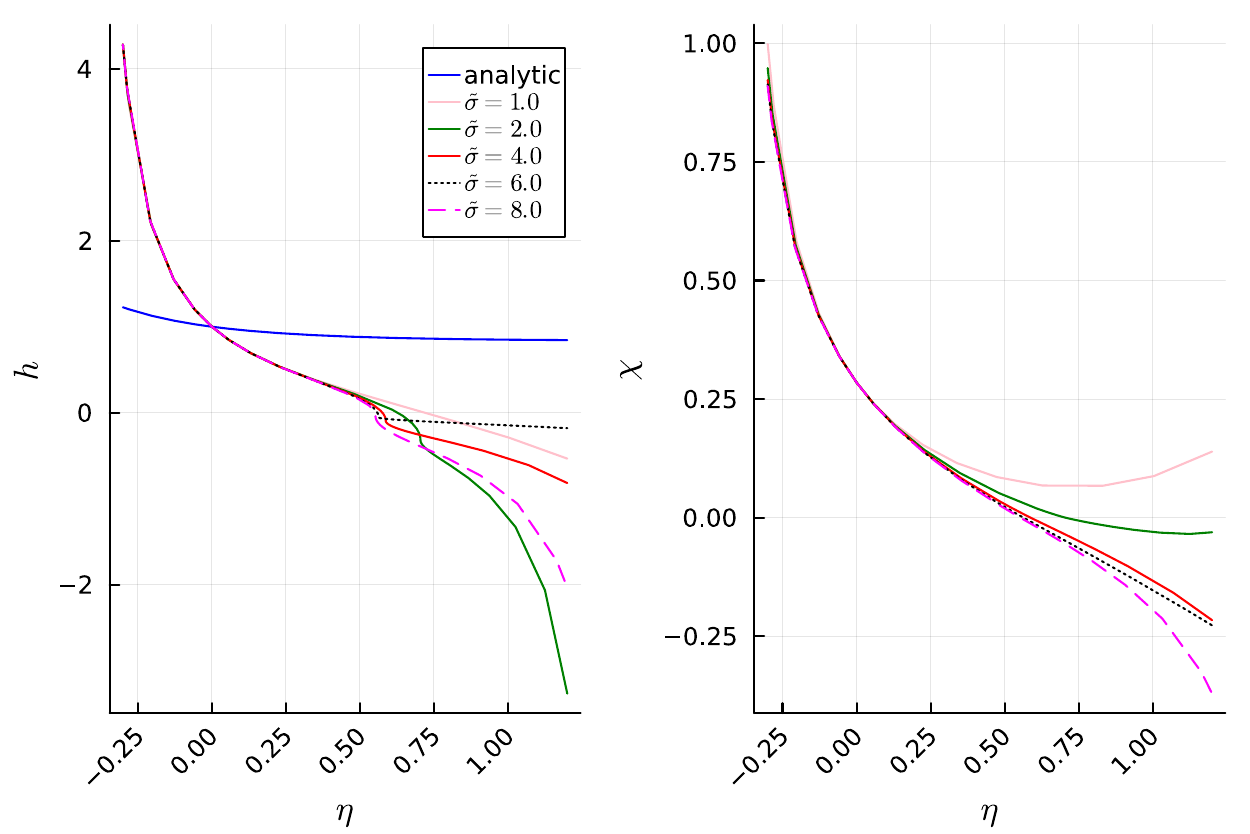}
\caption{The same plot as in \ref{span8} but form larger values of $\tilde{\sigma}$.}\label{span7}
\end{figure}

This is confirmed in Figure~\ref{span7} which shows the same functions,
but over a larger timescale. The differences between the models
become then visible at large $\eta$.

\begin{figure}[ht]
\centering
\includegraphics[scale=0.5]{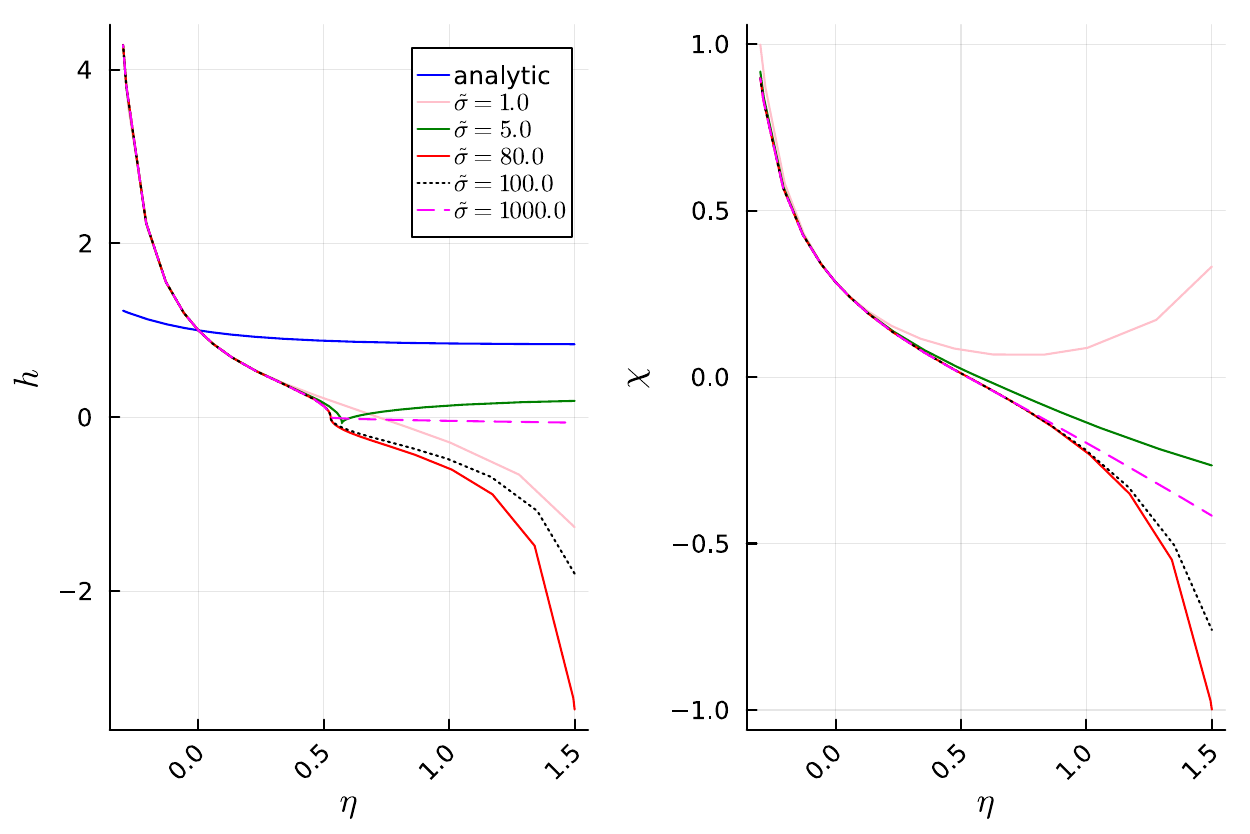}
\caption{The dimensionless Hubble function $h$ and the dimensionless
  density $\chi$ versus the dimensionless time $\eta$ for large values
  the parameter $\tilde{\sigma}$.}\label{span10}
\end{figure}

Increasing the values of $\tilde{\sigma}$ does not bring any new
insight, but for the first time there appears a cusp for
$\tilde{\sigma}=5$. MAPLE18 interpretes this as singularities. This is
demonstrated in Figure~\ref{span10}.

\begin{figure}[ht]
\centering
\includegraphics[scale=0.5]{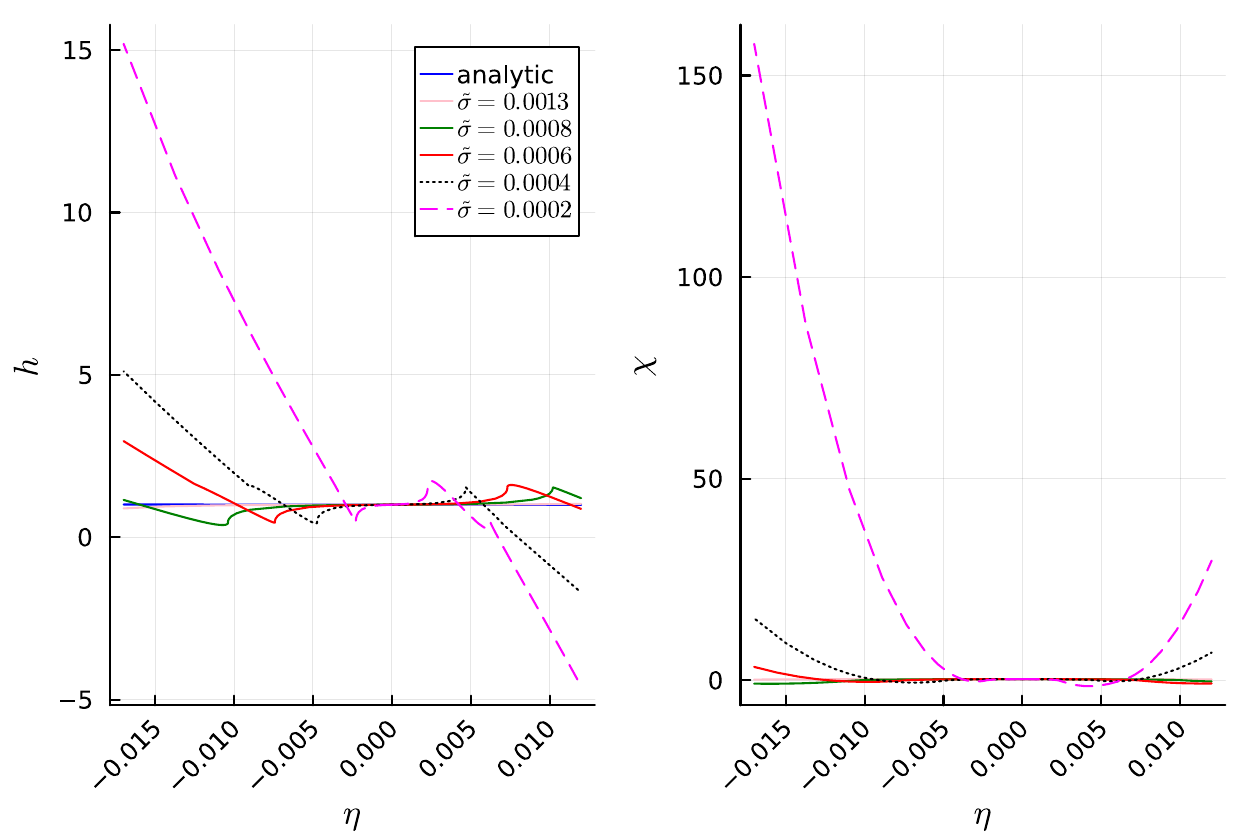}
\caption{The dimensionless Hubble function $h$ and the dimensionless
  density $\chi$ versus the dimensionless time $\eta$ for small values of
  the parameter $\tilde{\sigma}$.}\label{span11}
\end{figure}

In Figure~\ref{span11}, we return to relative small values of
$\tilde{\sigma}$ to show that the cusp behavior is quite common in the
model. This is interesting from a purely theoretical point of view, but
leaves doubts about the viability of the model over large time scales.

\begin{figure}[ht]
\centering
\includegraphics[scale=0.5]{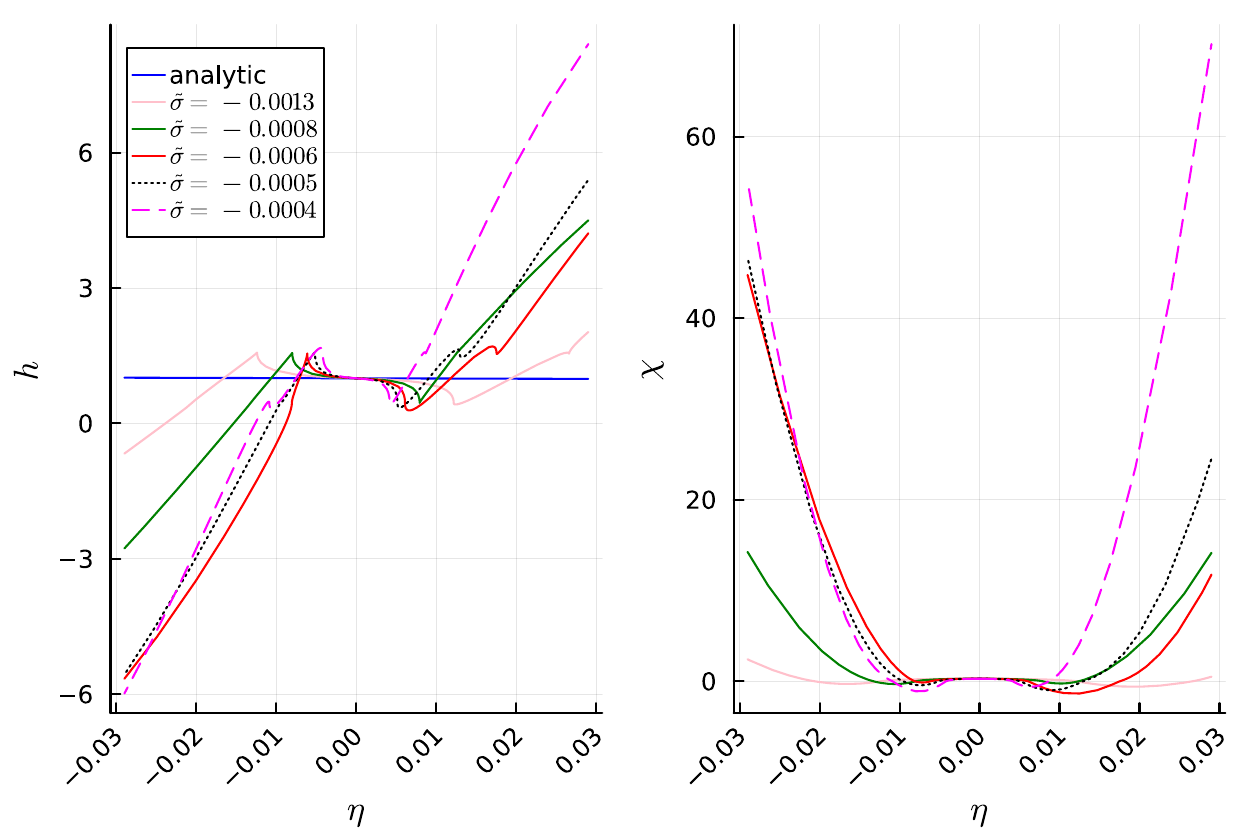}
\caption{The dimensionless Hubble function $h$ and the dimensionless
  density $\chi$ versus the dimensionless time $\eta$ for large negative
  values of
  the parameter $\tilde{\sigma}$.}\label{span13}
\end{figure}

\begin{figure}[ht]
\centering
\includegraphics[scale=0.5]{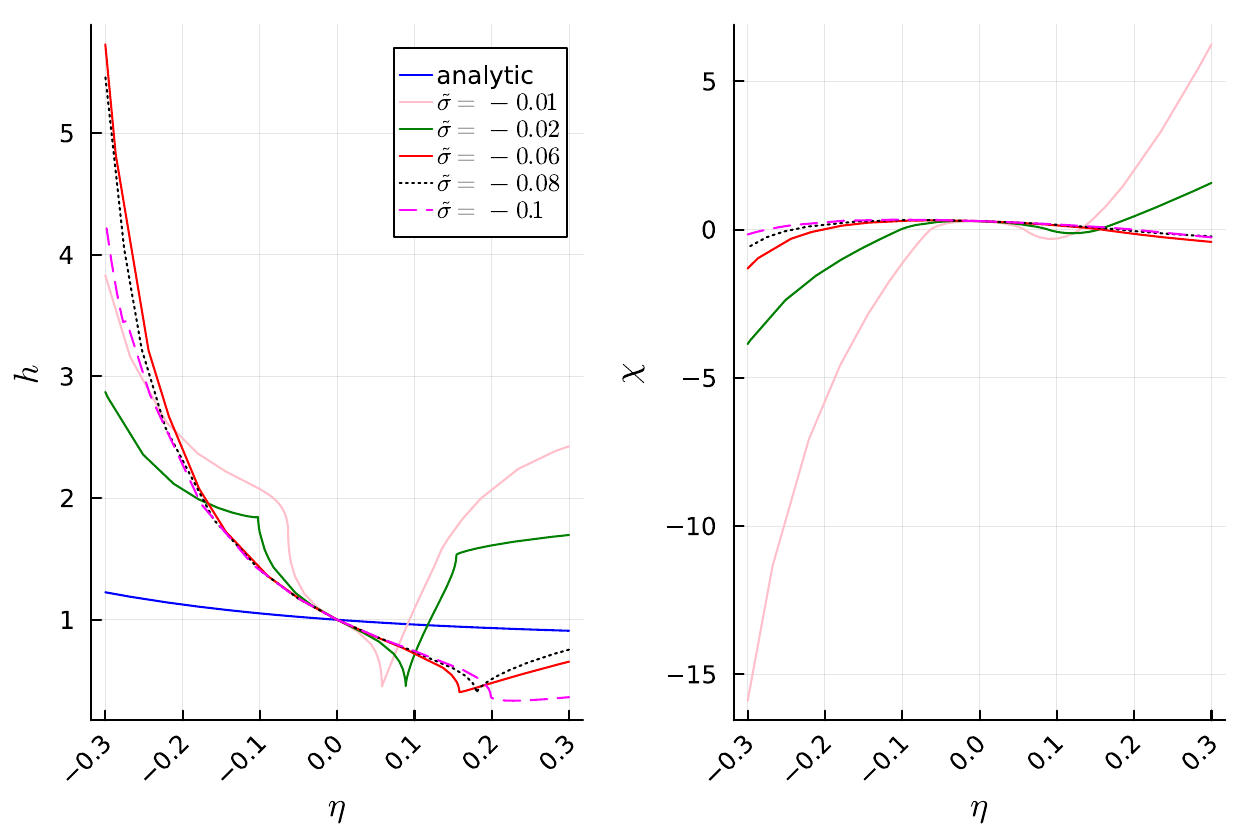}
\caption{The dimensionless Hubble function $h$ and the dimensionless
  density $\chi$ versus the dimensionless time $\eta$ for small negative
  values of
  the parameter $\tilde{\sigma}$.}\label{span15}
\end{figure}

\begin{figure}[ht]
\centering
\includegraphics[scale=0.5]{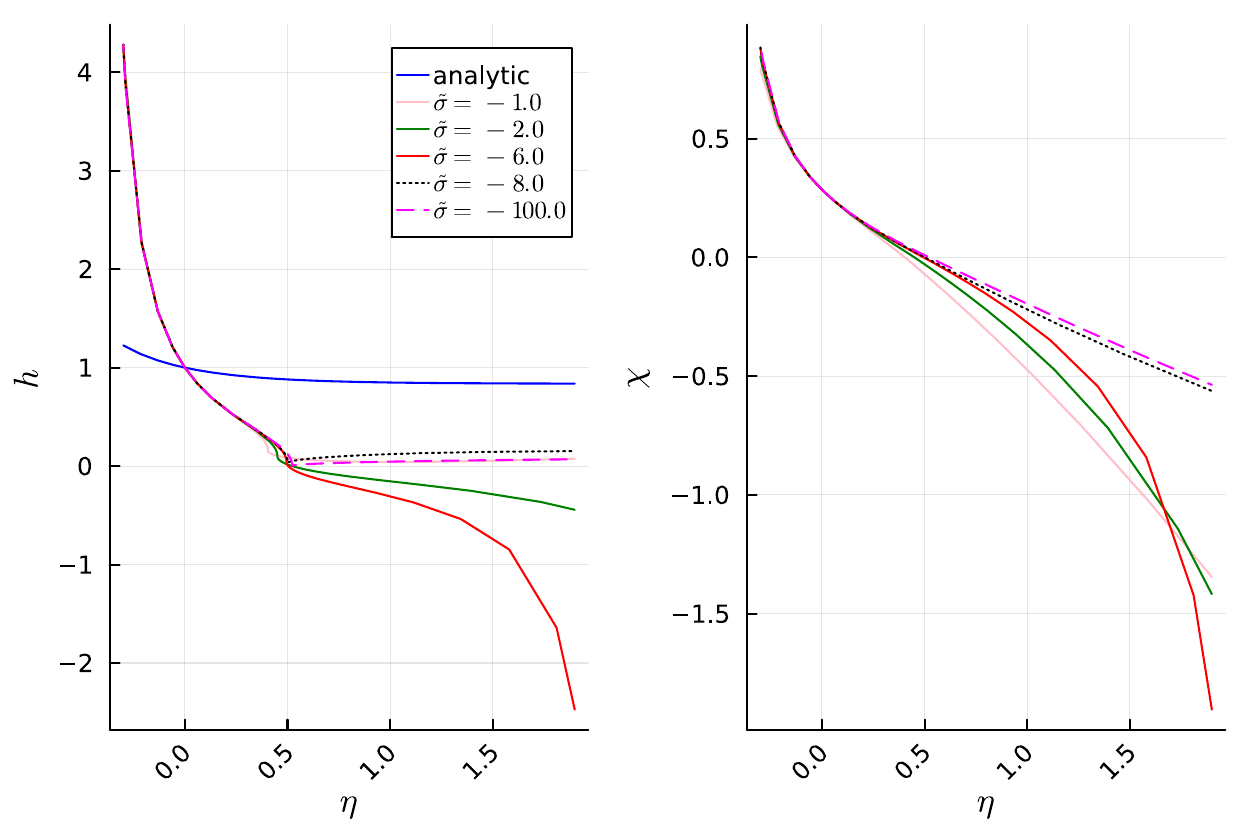}
\caption{The dimensionless Hubble function $h$ and the dimensionless
  density $\chi$ versus the dimensionless time $\eta$ for very small
  values of
  the parameter $\tilde{\sigma}$.}\label{span17}
\end{figure}

Finally, we turn our attention the the behaviour of $h$ and $\chi$ for
negative values of $\tilde{\sigma}$ which is depicted in Figures~
\ref{span13},~\ref{span15}, and \ref{span17}.

In Figure~\ref{span13}, we encounter a universe whose contraction
slows down to turn into a expanding universe at $\eta$ around
zero. Again many cusps accompany this behavior which could well mean
that universe undergoes a singularity there.

For small and very small values of negative $\tilde{\sigma}$ we have
again negative densities.

\begin{figure}[ht]
\centering
\includegraphics[scale=0.5]{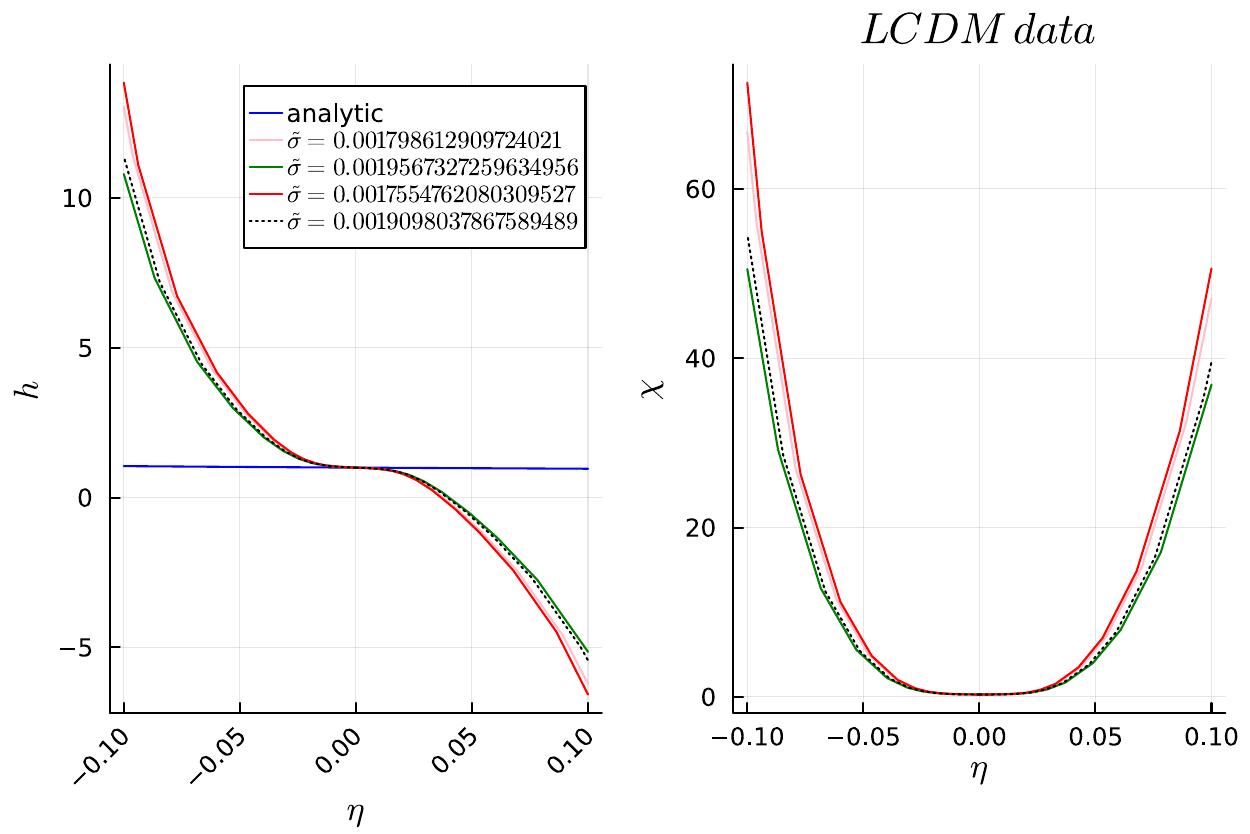}
\caption{The dimensionless Hubble function $h$ and the dimensionless
  density $\chi$ versus the dimensionless time $\eta$ for values of
  $\tilde{\sigma}$ resulting from equation (\ref{OmegaSigma}).}\label{LCDM}
\end{figure}

We have also explored a second approach for selecting the parameter
$\tilde{\sigma}$. More precisely, we utilized (\ref{OmegaSigma})
and considered the error bars associated with measured values in standard cosmology.
Figure~\ref{LCDM} provides an example where we calculated
$\tilde{\sigma}$ with high precision, staying within the range of
uncertainty. This allowed us to examine how sensitive the model is to the choice of parameters. The lifetime of the universe comes out
approximately
$|\eta_{univ}| \simeq 0.4$, which is too small to account for the
existence of the first galaxies in our universe. The Big
Crunch appears to be unavoidable again.

\begin{figure}[ht]
\centering
\includegraphics[scale=0.5]{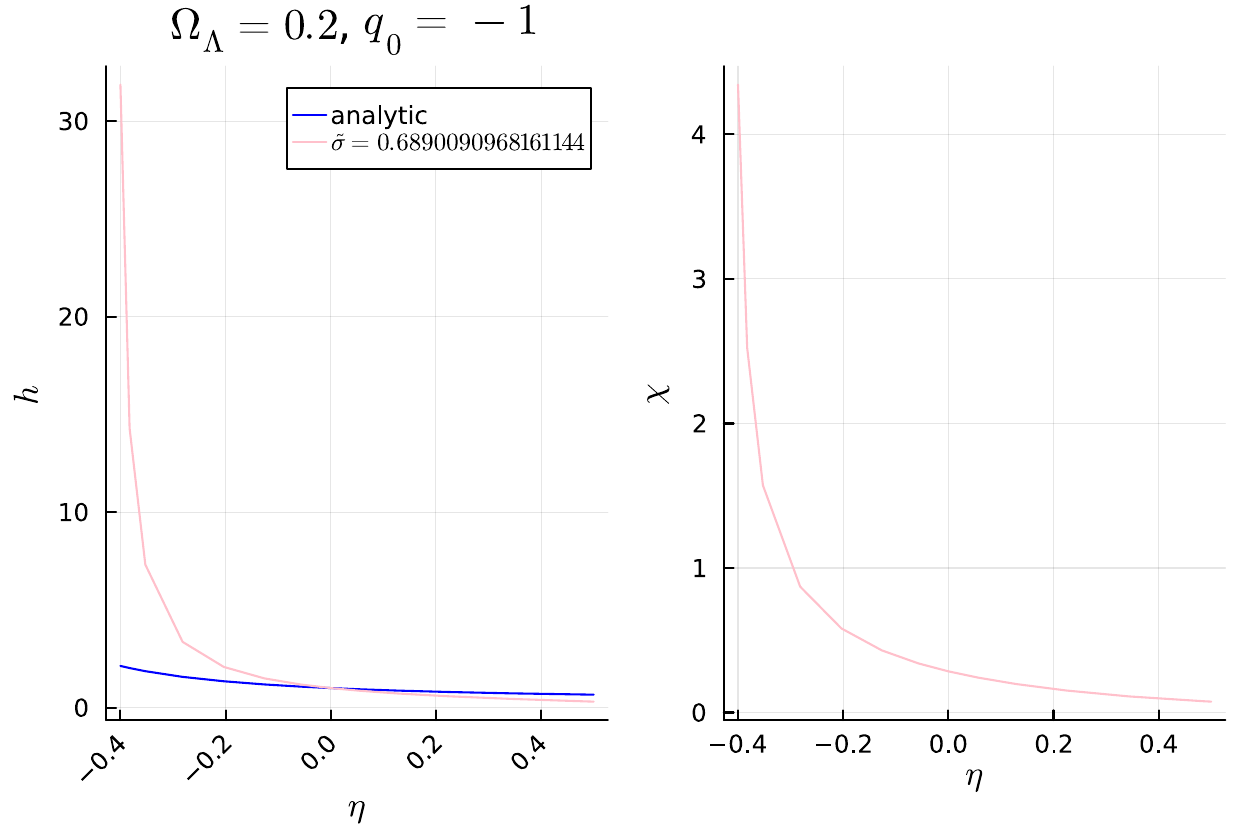}
\caption{The dimensionless Hubble function $h$ and the dimensionless
  density $\chi$ versus the dimensionless time $\eta$ using equation
  (\ref{Omega}) with values of the parameters as indicated in the Figure.}\label{lambda5}
\end{figure}

In the third approach, we used equation (\ref{Omega}) while keeping
the standard value for $\Omega_{m,0}$, but varying $\Omega_{\Lambda}$
and $q_0$. We then solved the equation to find $\tilde{\sigma}$. An
example is illustrated in Figure~\ref{lambda5}.
This variation led to a slight increase in the universe lifetime. Motivated by this observation, we
explored scenarios with a negative cosmological constant.  Two such
examples are noteworthy. Setting $q_0=-0.5$ and $\Omega_{\Lambda}=-50$ resulted in $\tilde{\sigma}\simeq 10$ and a lifetime of $|\eta_{univ}| \simeq 0.51$. Alternatively, when choosing a  deceleration parameter of $q_0=-1$, we obtained  $\tilde{\sigma} \simeq 7.8$ and a lifetime of  $0.56$. However, it appears challenging to significantly extend the model lifetime beyond these values.

In summary, this model represents a mild extension of
    Einstein gravity, particularly when viewed from the perspective of
    its Lagrangian. Surprisingly, the cosmology given by this model for a
spatially flat, homogenous universe is quite different from the
standard case based on Einstein equations.

\section{Conclusions}
Proposing a new gravity theory which goes beyond Einstein leads inevitably
to the examination of the new model in a cosmological context. In this context, theoretical advancements revealing new phenomena are equally important as empirical efforts to evaluate the model potential to supersede the standard cosmological framework.
We have emphasized both aspects studying the extension $f(R, {\cal
  L}_m)$ theory with an explicit coupling between geometry and matter
in the form $\sigma R \rho$ with $\rho$ being the energy density. This
appears as a natural choice used also in astrophysical context. We
explored the model over a wide range of the coupling constant $\sigma$
with initial data mimicking our present universe.  Given that the model predicts a universe lifespan that is drastically shorter than the observed age of our universe (approximately 13.8 billion years), it fails to meet a fundamental criterion for a viable cosmological model. This discrepancy leads us to conclude that the model, in its current form, cannot adequately describe the observed universe. As such, the potential occurrence of a Big Crunch, while theoretically interesting, becomes a secondary concern because the model is already ruled out based on its inability to account for the observed age of the universe.

All in all, the findings highlight
challenges in adapting the model to realistically represent the
universe, particularly regarding the universe lifetime and the issue
of negative densities in some scenarios.

Before comparing our work with existent literature, let us point out
that once we derive the field equations from the Lagrangian principle,
the Lagrangian should contain a matter part $L_m$ from which the
perfect energy-momentum tensor can be derived. The latter seems only
possible via a constrained variation with respect to the metric. The
conservation of the energy-momentum tensor in theories of the type
$f(R, L_m)$ is not guaranteed and requires special attention. In
\cite{Gonclaves} the choice of the Lagrangian is $f(R, L_m)=R/2+L_m
+\sigma RL_m$ with $L_m=-p$.  It is then difficult to study the dust
scenario with $p=0$ as it implies $L_m=0$. Furthermore, the
energy-momentum tensor is not conserved in this choice in contrast to
the authors' claim.  In \cite{ViscousDM} the choice is $f(R,L_m)=R/2
+L_m^{\alpha}$ with $\alpha$ being some number. It is not clear how we
can derive from that a hydrodynamical energy-momentum tensor and, as
explained above, whether we obtain a conservation law.  The choice of
the Lagrangian in \cite{ViscousDM} is also quite different from ours
as it lacks the direct coupling between geometry (represented by $R$)
and matter (represented by $L_m$). This shows that such coupling has
special consequences. In \cite{AcceleratingFRLm} the Lagrangian reads
$f(R,L_m)=R/2 +\alpha L_m^n - \beta$ with $\alpha$ and $\beta$ constants. Remarks similar
to the case of \cite{ViscousDM} would apply also here.



\end{document}